\documentclass[aps,pre,twocolumn,groupedaddress,amssymb,amsmath,showpacs,superscriptaddress]{revtex4}
\usepackage{graphicx}
\PassOptionsToPackage{caption=false}{subfig}
\usepackage{subfig}
\usepackage{color}

\newcommand{\beq}{\begin{equation}}
\newcommand{\eeq}{\end{equation}}
\newcommand{\beqn}{\begin{eqnarray}}
\newcommand{\eeqn}{\end{eqnarray}}

\begin{document}

\title{Extinction in neutrally stable stochastic Lotka-Volterra models}

\author{Alexander Dobrinevski}
\email[]{alexander.dobrinevski@lpt.ens.fr}
\affiliation{Arnold Sommerfeld Center for Theoretical Physics and Center for NanoScience, Department of Physics, Ludwig-Maximilians-Universit\"{a}t M\"{u}nchen, Theresienstrasse 37, D-80333 M\"{u}nchen, Germany}
\affiliation{CNRS-Laboratoire de Physique Th\'eorique de l'Ecole
Normale Sup\'erieure, 24 rue Lhomond, 75005 Paris
Cedex-France
\thanks{LPTENS is a Unit\'e Propre du C.N.R.S.
associ\'ee \`a l'Ecole Normale Sup\'erieure et \`a l'Universit\'e Paris Sud}
}

\author{Erwin Frey}
\email[]{frey@lmu.de}
\affiliation{Arnold Sommerfeld Center for Theoretical Physics and Center for NanoScience, Department of Physics, Ludwig-Maximilians-Universit\"{a}t M\"{u}nchen, Theresienstrasse 37, D-80333 M\"{u}nchen, Germany}

\begin{abstract}
Populations of competing biological species exhibit a fascinating interplay between the nonlinear dynamics of evolutionary selection forces and random fluctuations arising from the stochastic nature of the interactions. The processes leading to extinction of species, whose understanding is a key component in the study of evolution and biodiversity, are influenced by both of these factors. Here, we investigate a class of stochastic population dynamics models based on generalized Lotka-Volterra systems. In the case of neutral stability of the underlying deterministic model, the impact of intrinsic noise on the survival of species is dramatic: it destroys coexistence of interacting species on a time scale proportional to the population size. We introduce a new method based on stochastic averaging which allows one to understand this extinction process quantitatively by reduction to a lower-dimensional effective dynamics. This is performed analytically for two highly symmetrical models and can be generalized numerically to more complex situations. The extinction probability distributions and other quantities of interest we obtain show excellent agreement with simulations.
\end{abstract}


\maketitle

\section{Introduction\label{sec:Intro}}
Interactions between biological species are known to lead to very diverse and intricate behaviour of a population. This includes, just to give a few examples, coexistence of a surprisingly high number of competing species in the same ecological niche \cite{Hutchinson1961}, oscillating population cycles \cite{Gilg2003} and chaos \cite{Turchin2000}. The question which (if any) of the species in a web of interactions survive, and for how long, is thus very nontrivial but central for the understanding of evolution and biodiversity \cite{Frey2010}.

A classical and long-established model for the interaction of species in a well-mixed habitat is the Lotka-Volterra model \cite{Lotka1920,Volterra1931}. Since its introduction, it has also been successfully applied in many different contexts outside of population dynamics: among others, neural networks \cite{Rabinovich2008}, game theory \cite{Hofbauer1998} and physiology \cite{McCarley1975}. This model attempts to describe the interaction between $S$ species through a set of coupled ordinary differential equations of the form
\beq
\label{eq:GLVModel}
\partial_t x_i (t) = x_i (t) \left( b_i + \sum_{j=1}^{S} A_{ij} x_j (t) \right).
\eeq
The abundance of each species is given by a continuous, real-valued variable $x_i$ with $i=1,...,S$.
$b_i$ are constant source terms describing the growth (or decline) of each species in the absence of the others, and $A_{ij}$ is a constant matrix modelling the interactions between the species.
Within this model, survival or extinction of species is purely deterministic: Any fixed initial condition determines unambiguously which, if any, of the species survive. Technically, the main underpinning for this is the stability or instability of certain stationary solutions of the differential equations \eqref{eq:GLVModel}. Some rather precise criteria for determining the persistence of species directly from the vector $\vec{b}$, the matrix $A$ and the initial conditions have been obtained in literature \cite{Goh1977,Zeeman1995}.

However, in a real biological situation, the population is made up from a large but still finite number of individuals. Hence, the abundances of each species can only change in discrete steps and not continuously. Furthermore, the interactions between them, as well as birth and death processes, have -- to some extent -- a stochastic nature. All these features cannot be modelled by the deterministic equations \eqref{eq:GLVModel}.
In fact, such effects of finite system size and fluctuations due to some intrinsic randomness (or, likewise, due to external noise) have recently been recognized to be very important for extinction processes, especially in the case when the deterministic solutions exhibit neutral stability (see, for example, \cite{Reichenbach2006,Traulsen2006,Traulsen2005,Cremer2009,Nowak,Antal2006,Nowak2004a,Taylor2004,Fogel1998,Ficici2000} and many others).

In this paper, we propose a new method based on the idea of \textit{stochastic averaging} which allows one to gain a quantitative understanding of the stochastic extinction process in the case when the deterministic limit of the model is neutrally stable.  The idea of stochastic averaging was first introduced by Khasminskii in Ref. \cite{Khasminskii1968}. Later on, it was rigorously justified in Ref. \cite{Freidlin2004} for two-dimensional systems possessing a conservation law. So far, however, it has not gained a lot of popularity in physical literature.

In section \ref{sec:StochLV} we will formulate a stochastic model of population dynamics based on a graph of interactions between species, whose dynamics in the absence of noise reduces to a Lotka-Volterra model of Eq. \eqref{eq:GLVModel}.

In sections \ref{sec:RPS} and \ref{sec:4Sp} we will treat two pedagogical examples of such models, the three-species and four-species systems with cyclic dominance. Their deterministic dynamics will be shown to be neutrally stable, i.e. to lead to perpetual coexistence of all species with periodically oscillating abundances. However, we will see that taking into account fluctuations due to the stochastic nature of the interactions introduces a finite mean extinction time proportional to the population size. Using stochastic averaging, we will characterize the extinction process by an effective stochastic process in the deterministically conserved quantities. This will allow us to obtain quantitative results on extinction times and their dependence on the initial conditions.

The generalization to more complex models will be discussed in section \ref{sec:Conclusion}.

\section{Stochastic Lotka-Volterra Models\label{sec:StochLV}}

\begin{figure}[t]
\begin{center}
\includegraphics[width=0.7\columnwidth]{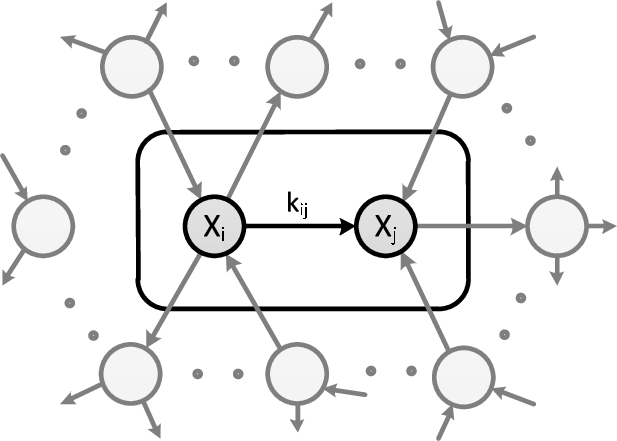}
\end{center}
\caption{Subgraph of a complex interaction graph.}
\label{fig:GLVGraph}
\end{figure}

Let us consider a well-mixed population with $S$ species, and interactions between them defined by a graph $G=(V,E)$ with vertices $V=\{X_1,...,X_S\}$ and  (directed) edges 
\[E=\{(X_i,X_j)|i \neq j;\, i,j=1,...,S\}.\]
An edge from $X_i$ to $X_j$ is denoted by an arrow in figure \ref{fig:GLVGraph}, and is taken to indicate that the species $X_i$ dominates over $X_j$. We allow at most one edge between each pair of species.
From this graph, we can now write down a set of reaction equations implementing the interactions of the species. For every edge $(X_i,X_j)\in E$ we formulate an interaction between $X_i$ and $X_j$ in the formalism of chemical reaction equations:
\beq
\label{eq:GLVReactionEqns}
X_i+X_j \stackrel{k_{ij}}{\longrightarrow} X_i+X_i,
\eeq
where the reaction rate $k_{ij}$ is the probability for this reaction to occur per (infinitesimal) time unit $dt$ and per possible pair of individuals.

Note that the model described by the reactions in Eq. \eqref{eq:GLVReactionEqns} provides an individual-based, discrete description of the interactions, and includes stochastic fluctuations since the reaction rates are interpreted probabilitistically.

Since the reactions in Eq. \eqref{eq:GLVReactionEqns} keep the total number of individuals fixed, we can assume a constant system size $N$. The system state is then described by the $S$-tuple of species counts $\vec{n}=\left(n_1,...,n_S\right)$, with $n_i \in {0,...,N}$ and $\sum{n_i}=N$. We define the $S$ ``basis vectors'' $\vec{e}_1,...,\vec{e}_S$ by $\vec{e}_i=(0,...,0,1,0,...,0)$ where the $1$ is on the $i$-th position.
The reactions in Eq. \eqref{eq:GLVReactionEqns} can then be translated into a master equation giving the evolution of the occupation probabilities $P_{\vec{n}}(t)$ for each state $\vec{n}$:
\beqn
\nonumber
\partial_t P_{\vec{n}}(t) &= \sum\limits_{(X_i,X_j) \in E} k_{ij} &\left[(n_i-1)(n_j+1)P_{\vec{n}-\vec{e}_i+\vec{e}_j}(t) \right. \\
& & \left.-n_i n_j P_{\vec{n}}(t) \right].
\label{eq:GLVMasterEqn}
\eeqn

For biological applications, one is mostly interested in large populations, i.e. in the limit of large $N$. The relative abundances of each species, $x_k=\frac{n_k}{N}$, can then be assumed to be real-valued variables in the interval $[0;1]$. Using  a standard Kramers-Moyal expansion \cite{VanKampen}, the master equation \eqref{eq:GLVMasterEqn} can then be approximated by a Fokker-Planck equation in the intensive variables $x_k$:
\beqn
\nonumber
\partial_t P(\{x_k\},t) &=& -\sum_{i=1}^S \partial_i\left[ \alpha_i P(\{x_k\},t)\right] \\
& & +\frac{1}{2N}\sum_{i,j=1}^S \partial_i\partial_j \left[B_{ij} P(\{x_k\},t)\right].
\label{eq:GLVFPE}
\eeqn
The conservation of the total population size $N$ gives rise to the normalization condition $\sum{x_i}=1$. The drift and noise terms in Eq. \eqref{eq:GLVFPE} are given by:
\beqn
\label{eq:GLVFPEAlpha}
\alpha_i & = & x_i \sum_{j=1}^S A_{ij} x_j, \\
\label{eq:GLVFPEB}
B_{ij} & = & \left\{ \begin{array}{ccc}
	\sum_{k=1}^S \left|A_{ik}\right| x_i x_k & \mathrm{for} & i=j \\
	-\left|A_{ij}\right| x_i x_j & \mathrm{for} & i \neq j 
\end{array} \right. .
\eeqn

The entries of the interaction matrix $A$ are:
\beq
\label{eq:GLVIntMat}
A_{ij} = \left\{ 
\begin{array}{ccc} k_{ij} &&\quad\mbox{if}\quad (X_i,X_j) \in E \\
-k_{ij} &&\quad\mbox{if}\quad (X_j,X_i) \in E \\
0 &&\quad\mbox{otherwise}
\end{array} \right. .
\eeq

As is well known \cite{Gardiner}, the Fokker-Planck equation \eqref{eq:GLVFPE} can be reformulated as a set of  stochastic differential equations, or \textit{Langevin equations}: 
\beq
\label{eq:GLVSDE}
dx_i = \alpha_i dt + \frac{1}{\sqrt{N}} \sum_{j=1}^S \mathcal{C}_{ij}\,dW_j .
\eeq
Throughout this paper, we take all stochastic differential equations to be in the It\^{o} interpretation.
$\mathcal{C}$ is a matrix satisfying $\mathcal{C}\mathcal{C}^T=B$, with $B$ defined by Eq. \eqref{eq:GLVFPEB}. Certainly, this does not fix $\mathcal{C}$ uniquely, but the precise choice has no influence on the stochastic process \cite{Gardiner}. $W_j$ are independent Wiener processes, or Brownian motions, with zero mean and unit variance.

From equation \eqref{eq:GLVSDE} we see that the deterministic, noiseless limit of the general model of Eq. \eqref{eq:GLVReactionEqns} is given by the following set of coupled ordinary differential equations (so-called \textit{rate equations}):
\beq
\label{eq:GLVRateEquations}
\partial_t x_i = x_i  \sum_{j=1}^{S} A_{ij} x_j \, .
\eeq
This can now be immediately identified as a generalized Lotka-Volterra model of the form of Eq.\eqref{eq:GLVModel}. The interaction matrix $A$ is given by Eq. \eqref{eq:GLVIntMat}, and the source terms vanish, i.e. $b_i=0$, simply since all reactions in Eq. \eqref{eq:GLVReactionEqns} have exactly two reactants. A model with non-vanishing source terms can be built by considering reactions of the form of death processes, $X_i\rightarrow \emptyset$, and birth (branching) processes, $X_i\rightarrow 2X_i$, in addition to Eq. \eqref{eq:GLVReactionEqns}.

The noise terms in Eq. \eqref{eq:GLVSDE}, proportional to $\frac{1}{\sqrt{N}}$, encapsulate the fluctuations due to the discreteness of the individuals and the stochastic nature of the reactions in Eq. \eqref{eq:GLVReactionEqns}.

In total, we have given a procedure allowing us to obtain a stochastic model (in terms of Fokker-Planck or Langevin equations) of a system with a large population of individuals from a general interaction graph of the species. This prescription is certainly not unique, however, it has the nice property that the deterministic dynamics of the resulting model is in one-to-one correspondence with a Lotka-Volterra model whose interaction matrix is the adjacency matrix of the original graph.

Considering the widespread use and the importance of Lotka-Volterra models, it seems worthwhile to study models of the form of Eq. \eqref{eq:GLVSDE} and the effects of stochasticity in them.

In general, the deterministic rate equations \eqref{eq:GLVRateEquations} possess extinction fixed points (where some species are extinct, i.e. there are some $j$ with $x_j=0$) and coexistence fixed points (where all species are present, $x_i > 0$ for all $i$). From Eq. \eqref{eq:GLVRateEquations} we see immediately that the coexistence fixed points form a linear subspace of the phase space of the system which is given by the kernel of the matrix $A$.
When the full stochastic model in Eq. \eqref{eq:GLVSDE} is considered, fluctuations cause the system to touch an extinction hyperplane where $x_i = 0$ for some $i$ sooner or later. Since a species which has died out cannot be re-introduced (this is apparent from the reaction equations \eqref{eq:GLVReactionEqns}), this means that in the stochastic system, \textit{extinction always occurs eventually}. 

However, the time scale on which this process occurs can vary greatly. A classification of the possible scenarios, characterized by the scaling of the mean extinction time $T_{\mathrm{ext}}$ with the population size $N$ was discussed in Ref. \cite{Antal2006} and further developed in Ref. \cite{Reichenbach2007, Cremer2009}:\begin{enumerate}
	\item \textit{Stable} coexistence for $T_{\mathrm{ext}}\propto e^N$, occurring when the deterministic dynamics has a stable attractor in the coexistence region. Here, extinction is driven by rare large deviations and hence the extinction times for large populations are extremely long.
	\item \textit{Unstable} coexistence for $T_{\mathrm{ext}}\propto \log{N}$. This occurs when the flow of the deterministic dynamics approaches one of the extinction hyperplanes for large times, and weak fluctuations are already sufficient to make one of the species go extinct.
	\item \textit{Neutrally stable} coexistence for a power-law dependence, $T_{\mathrm{ext}}\propto N^{\gamma}$. This occurs when the deterministic dynamics possesses a family of neutrally stable, closed orbits, corresponding to the existence of a conservation law. 
\end{enumerate}
For simple models, these criteria correspond to (linearly) stable, unstable, or neutrally stable coexistence fixed points in the rate equations \eqref{eq:GLVReactionEqns}. 

While here we are only considering well-mixed populations, a similar classification is possible for models which include spatial degrees of freedom \cite{Reichenbach2007}. For a review on population dynamics in spatially extended systems see e.g. \cite{Szabo2007,Frey2010}. There is yet another interesting connection to extinction times close to absorbing-state phase transitions; see e.g. Refs.~\cite{Odor2004,Henkel2009,Kuhr2011}.

Observe that the effects of stochasticity are most dramatic in a neutrally stable model: while the deterministic dynamics predicts perpetual coexistence far away from the extinction planes, inclusion of fluctuations introduces a finite mean extinction time which only scales { as a power law} with the population size.
In the following two sections, we will now analyze the stochastic extinction process for two pedagogical example models of this kind. 

\section{Cyclic Three-Species Model: The Rock-Paper-Scissors Game\label{sec:RPS}}

\subsection{Introduction}

\begin{figure}[t]
\centering
\includegraphics[width=0.4\columnwidth]{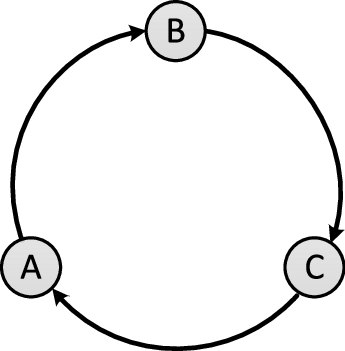}
\caption{Interaction graph for the rock-paper-scissors game}
\label{fig:3SpeciesCyclic}
\end{figure}

The first example that shall be considered in detail is a three-species model with cyclic dominance, whose interaction graph is shown in figure \ref{fig:3SpeciesCyclic}. One of the most popular areas where such cyclic, intransitive relationships between three entities arise is in game theory as a so-called \textit{rock-paper-scissors game} \cite{Hofbauer1998}. In a more biological context, they have been observed between strains of \textit{E. coli} bacteria \cite{Kerr2002} and between lizard morphs \cite{Sinervo1996}. Another rather different application is to forest fire models \cite{Clar1996}, where the three states trees, fire and ash obey a similar relatioship.

The reaction equations for this model, according to the general treatment in section \ref{sec:StochLV}, read:
\beqn
\nonumber A+B & \rightarrow A+A , \\
\nonumber B+C & \rightarrow B+B , \\
\label{eq:RPSReactionEqns} C+A & \rightarrow C+C .
\eeqn
The interaction matrix is, accordingly:
\beq
\label{eq:RPSFPEAdj}
(A_{ij})=\left(\begin{array}{ccc}
	0 & 1 & - 1 \\	-1 & 0 & 1 \\ 1 & -1 & 0
\end{array}\right) .
\eeq

In order to simplify the calculations, we have set all reaction rates to be equal in Eq. \eqref{eq:RPSFPEAdj}. By rescaling time we can then set them to $1$ without loss of generality. According to the general treatment in section \ref{sec:StochLV}, the stochastic model in the large-$N$ limit is then described by the Fokker-Planck equation \eqref{eq:GLVFPE} (or, equivalently, the Langevin equations \eqref{eq:GLVSDE}) with $S=3$. The drift term in Eq. \eqref{eq:GLVFPEAlpha} and the noise term in Eq. \eqref{eq:GLVFPEB} evaluate explicitely to the following expressions:
\beqn
\label{eq:RPSFPEAlpha}
\vec{\alpha} &= & \left(\begin{array}{c}
	a( b- c) \\
	b( c- a) \\
	c( a- b)
\end{array}\right) , \\
\label{eq:RPSFPEB}
B & = & \left(\begin{array}{ccc}
	a( b+ c) & - a b & - a c \\
	- a b & b( c+ a) & - b c \\
	- a c & - b c & c( a+ b)
\end{array}\right) .
\eeqn
Note that here and in the following we shall use the variable names $a$, $b$, $c$ and $x_1$, $x_2$, $x_3$ interchangeably.
A qualitative treatment of precisely this model was given in Ref. \cite{Reichenbach2006}. In the following, we shall briefly summarize the previous results relevant for our considerations.

\begin{figure}[t]
\centering
\includegraphics[width=0.7\columnwidth]{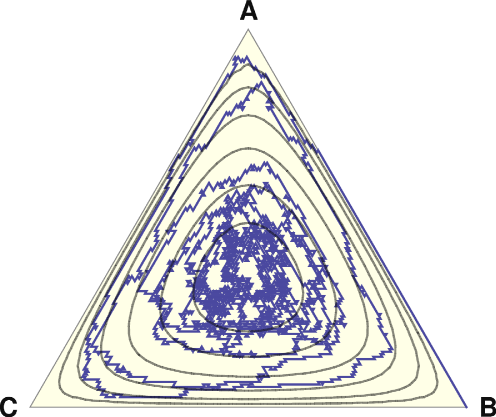}%
\caption{{(Color online)} Phase space of the three-species rock-paper-scissors game. Grey curves: closed deterministic orbits, given by $\rho=const.$. Blue (or dark grey) curve: example of a stochastic trajectory, obtained for a system size of $N=200$.}%
\label{fig:PhaseSpaceRPS}%
\end{figure}

The deterministic model, obtained by dropping the noise terms from Eq. \eqref{eq:GLVSDE}, is given by the rate equations $\partial_t x_i = \alpha_i$ with $\vec{\alpha}$ as in Eq. \eqref{eq:RPSFPEAlpha}. Due to the normalization condition $a+b+c=1$, its phase space can be viewed as the $2$-simplex, i.e. an equilateral triangle. Its corners correspond to complete extinction (i.e. only one species of the three species is present), and its edges to states where one species is extinct and two are still present. The dynamics of the rate equations yields oscillations along closed, periodic orbits around a coexistence fixed point at $a=b=c=\frac{1}{3}$. Close to the fixed point, the orbits are almost circular, whereas further away, they approach the triangular shape of the simplex boundaries (see figure \ref{fig:PhaseSpaceRPS}). These orbits are neutrally stable due to the existence of a conserved quantity 
\beq
\rho = abc.
\eeq
$\rho$ assumes its maximum value $\rho_{\mathrm{max}}=\frac{1}{27}$ at the coexistence fixed point in the center of the phase space triangle, and its minimum value $\rho=0$ on the edges, corresponding to extinction of at least one species.

In Ref. \cite{Reichenbach2006}, it was also shown that with the inclusion of noise in the full stochastic model in Eq. \eqref{eq:GLVFPE}, the evolution of the population no longer takes place deterministically along one closed orbit, but can fluctuate randomly between different orbits, cf. Fig.~\ref{fig:PhaseSpaceRPS}. By means of a linearization around the coexistence fixed point, it was derived that eventually, the stochastic trajectory will hit one of the simplex boundaries, and from there move towards one of the absorbing corners of the triangle. This process means that two of the three species go extinct when stochasticity is included. It was also motivated that the mean extinction time scales as $T_{\mathrm{ext}}\propto N$.

Here, instead of linearizing the stochastic model in Eq. \eqref{eq:GLVFPE}, we will perform a stochastic averaging procedure over the deterministic orbits. This will remove the fast, oscillatory degrees of freedom (taking into account all non-linearities and the precise geometry of the phase space) and produce an effective one-dimensional stochastic differential equation for $\rho$. Through this, we will obtain an exact description of the extinction process and quantitatively correct results for mean extinction times.

\subsection{Stochastic Averaging\label{sec:RPSStochAvg}}

Let us start with the formulation of the stochastic model using the It\^{o} stochastic differential equations \eqref{eq:GLVSDE}. Since the deterministic drift terms in Eq. \eqref{eq:GLVSDE} keep $\rho=abc$ conserved, this quantity changes only due to the noise terms $\propto \frac{1}{\sqrt{N}}$, i.e. much more slowly than the oscillations along an orbit with constant $\rho$. Furthermore, $\rho$ is a measure for closeness to extinction in the sense that the time when $\rho$ becomes $0$ for the first time is exactly the time when the first of the three species goes extinct. Thus, a description in terms of $\rho$ allows us to separate the deterministic dynamics (i.e. the rapid oscillations along the closed orbits), which does not contribute to extinction, from the stochastic fluctuations which lead to movement between different orbits and ultimately cause one of the species to die out.

To determine the dynamics of $\rho$ quantitatively, we use Eq. \eqref{eq:GLVSDE} and apply the It\^{o} chain rule \cite{Gardiner}, giving:
\beqn
\nonumber
d\rho &=& \left( \sum_{i=1}^3\alpha_i\partial_i\rho + \frac{1}{2N}\sum_{i,j=1}^3 B_{ij}\partial_i\partial_j\rho \right)dt \\
& & + \frac{1}{\sqrt{N}}\sum_{i,j=1}^3\left(\mathcal{C}_{ij}\partial_i\rho\right)\, dW_j .
\label{eq:RPSRhoEqn1}
\eeqn
The first term $\sum_i\alpha_i\partial_i\rho$ is zero since $\rho$ is conserved by the rate equations.
Eq. \eqref{eq:RPSRhoEqn1} then implies that { $\dot{\rho} \sim \frac{1}{N}$, i.e. that $\rho$ changes on a slow time scale $\propto N$ and that coexistence in our model is neutrally stable. Actually, there is a more general relationship between the existence of conserved quantities and neutral stability of species \footnote{{Any quantity $f(\vec{x})$ conserved by the rate equations \eqref{eq:GLVRateEquations} in a system of $S$ species with arbitrary reaction rates also satisfies $\sum_{i=1}^S \alpha_i\partial_i f(\vec{x})=0$. Thus, as in Eq.~\eqref{eq:RPSRhoEqn1}, the evolution of such a conserved quantity $f(\vec{x})$ in the stochastic system occurs on a slow time scale $\sim N$. Suppose now there exists a conserved quantity which is a non-trivial function of the abundances of a subset of species $F=\{s_1,...,s_M\} \subset \{1,...,S\}$, and that $f(\vec{x})= f^*$ denotes a manifold in phase space where at least one of the species in the subset $F$ goes extinct. Starting from an initial condition $\vec{x}_0$ with $f(\vec{x}_0)$ distinct from $f^*$, the time to reach this manifold has to scale $\sim N$. This implies that the coexistence of the species in subset $F$ is neutrally stable in the classification scheme of section \ref{sec:StochLV}.}}.}

The second term in Eq. \eqref{eq:RPSRhoEqn1} is a ``stochastic drift'' term arising from the fluctuations, and evaluates to:
\beq
\frac{1}{2N}\sum_{i,j=1}^3 B_{ij}\partial_i\partial_j\rho = -\frac{3}{N}\rho .
\eeq
Khasminskii's stochastic averaging theorem \cite{Khasminskii1968} now states that {to leading order in $\frac{1}{N}$, the evolution of $\rho$ is exactly described} by the stochastic differential equation
\beq
\label{eq:RPSSDEEffRho}
d\rho = -\frac{3}{N}\rho\, dt + \frac{1}{\sqrt{N}} \sqrt{\overline{D}(\rho)}\,dV ,
\eeq
{ where $V$ is a Wiener process with zero mean and unit variance. $\overline{D}(\rho)$ is} an averaged diffusion coefficient given by 
\beqn
\label{eq:RPSSDEEffRhoD}
\overline{D}(\rho)&:=&\frac{1}{T(\rho)}\int_0^{T(\rho)}{D(a(t),b(t),c(t))\,dt} , \\
\nonumber
D(a,b,c) &=& \left[\mathcal{C}^T(\nabla\rho)\right]^T\left[\mathcal{C}^T(\nabla\rho)\right]
\\
\nonumber
&=& (\nabla\rho)^T B(\nabla\rho) = \rho^2 \left( -9 + \frac{1}{a} + \frac{1}{b} + \frac{1}{c} \right) .
\eeqn
{Note that $D(a,b,c) = \sum_i\left[\sum_{j=1}^3\mathcal{C}_{ij}\partial_i\rho\right]^2$ is the total variance of the noise terms in the Langevin equation for $\rho$, Eq.~\eqref{eq:RPSRhoEqn1}. Khasminskii's theorem thus tells us that due to time scale separation, on the slow time scale $t\sim N$ these noise terms may be treated as independent and replaced by a single effective noise source~\footnote{ Khasminskii's work \cite{Khasminskii1968} actually applies to
general Markovian processes with an arbitrary number of slow and fast
variables. It shows that on a fixed time interval, when the time scale
separation becomes stronger the stochastic process in the slow variables of
the full model converges (in some weak sense) to the solution of the
stochastically averaged SDE for the slow variables. We refer the interested
reader to Ref. \cite{Khasminskii1968} for details on the mathematical
formulation, including the precise continuity requirements on the drift and
noise terms and the precise statement of the convergence results.}. Its variance, given in Eq. \eqref{eq:RPSSDEEffRhoD}, is the }
time-average of $D(a,b,c)$ over the closed orbit of the deterministic rate equations corresponding to a fixed value of $\rho$. { $T(\rho)$ is the period of this orbit, and} $a(t)$, $b(t)$, $c(t)$ parametrize this deterministic orbit in terms of the time $t$. 

Note that Eq. \eqref{eq:RPSRhoEqn1} is not a closed equation. It describes the dynamics of $\rho$ only as a function depending on the dynamics of the variables $a$, $b$, $c$, since the prefactors of the noise terms depended on all three variables individually. In contrast, Eq. \eqref{eq:RPSSDEEffRho} is a closed equation uniquely defining a stochastic process $\rho(t)$, now viewed as a stochastic variable evolving on the interval $\left[ 0;\rho_{\mathrm{max}}=\frac{1}{27}\right]$.

Intuitively, this averaging procedure is justified by the time scale separation described above: The deterministic drift along the orbit with constant $\rho$ takes place on a time scale of $\mathcal{O}(1)$ fixed by the rate equations, and the movement between different orbits -- i.e. the changes in $\rho$ due to the noise terms in the stochastic differential equations \eqref{eq:GLVSDE} -- occur on a time scale of $\mathcal{O}(N)$. This allows one to average over the fast deterministic evolution and noise in the phase variable, leaving an effective slow process given by Eq. \eqref{eq:RPSSDEEffRho}.

Eq. \eqref{eq:RPSSDEEffRho} can be reformulated equivalently as a Fokker-Planck equation for the probability distribution $P(\rho,t)$:
\beq
\label{eq:RPSFPEEffRho}
\partial_t P(\rho,t) = \frac{1}{N}\left\{ \partial_\rho \left[3\rho\,P(\rho,t)\right] + \frac{1}{2} \partial_\rho^2 \left[\overline{D}(\rho)\,P(\rho,t) \right] \right\} .
\eeq
In this form it is most apparent that the time scale of the extinction process is $t\propto N$.

\begin{figure}[t]
\centering
\includegraphics[height=0.65\columnwidth]{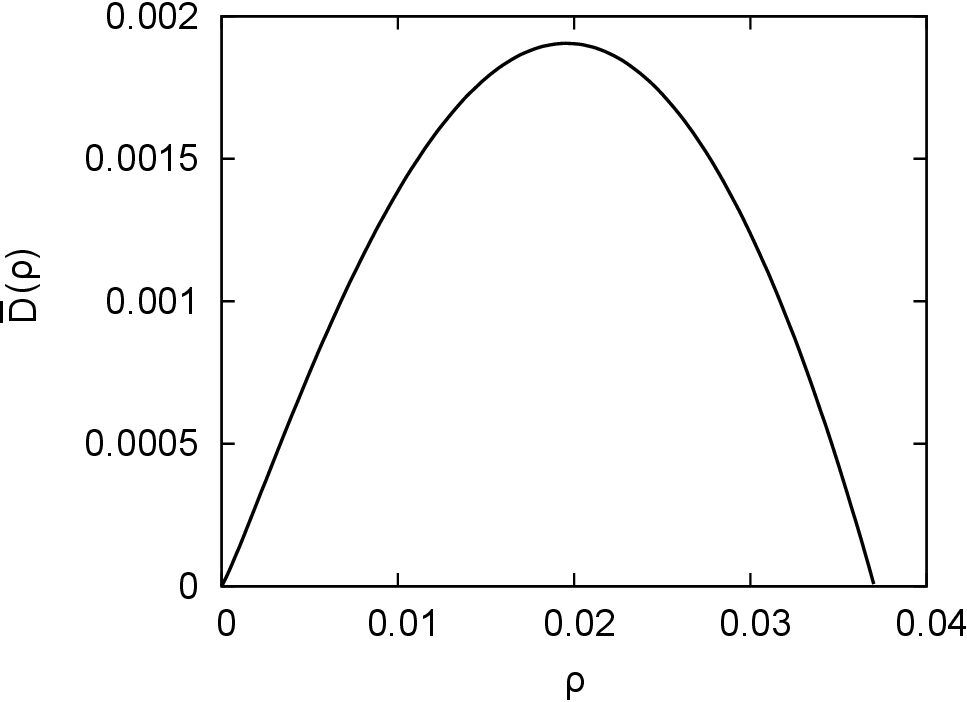}%
\caption{Effective diffusion coefficient $\overline{D}(\rho)$ as given in Eq. \eqref{eq:RPSDiffCoeff}. Observe that $\overline{D}(\rho)$ vanishes both at $\rho=0$ (extinction boundary of phase space triangle) and at $\rho=\rho_{\mathrm{max}}=\frac{1}{27}$ (coexistence fixed point in the center of the phase space triangle).}%
\label{fig:RPSDeff}%
\end{figure}

In this simple model with its high degree of symmetry, the integral in Eq. \eqref{eq:RPSSDEEffRhoD} can be performed analytically to give a closed expression for $\overline{D}(\rho)$ in terms of $\rho$. A sketch of the computation is given in appendix \ref{sec:AppCalc3Sp}, and the resulting formula is:
\beq
\label{eq:RPSDiffCoeff}
\overline{D}(\rho) = 3\rho^2 \left[ -3 + \frac{1}{a_1} +\left( \frac{1}{a_{\mathrm{min}}} - \frac{1}{a_{1}} \right) \frac{E(k)}{K(k)} \right] .
\eeq
Here, $K(k)$ and $E(k)$ are complete elliptic integrals of the first and second kind, respectively. The elliptic modulus is given by 
\beq
\label{eq:RPSEqRatesModulus}
k^2=\frac{(a_{\mathrm{max}}-a_{\mathrm{min}})a_1}{(a_1-a_{\mathrm{min}})a_{\mathrm{max}}},
\eeq
and $a_{\mathrm{min}}$, $a_{\mathrm{max}}$ and $a_1$ are the three real roots of the polynomial 
\beq
a\left( 1-a \right)^2 = 4\rho.
\eeq
As is well-known, these roots can be written down explicitely in terms of $\rho$. Graphically, $\overline{D}(\rho)$ is shown in Fig.~\ref{fig:RPSDeff}. { At the boundary $\rho=0$, i.e. near extinction, one obtains the asymptotic form
\beq
\nonumber
\overline{D}(\rho) = -\frac{\rho}{\ln \rho} + \mathcal{O}\left(\rho^{2}, \frac{\rho^2}{\ln \rho}, \frac{\rho^2}{(\ln \rho)^2}\right).
\eeq
On the other hand, at the boundary $\rho = \frac{1}{27}$, i.e. with nearly equal concentrations of the three species, one obtains the asymptotics
\beq
\label{eq:DEffCoex}
\overline{D}(\rho) = \frac{2}{9}\left(\frac{1}{27}-\rho\right) + \mathcal{O}\left(\frac{1}{27}-\rho\right)^2.
\eeq
We can now make contact with the results of Ref. \cite{Reichenbach2006}, which were obtained by linearizing in the vicinity of the coexistence fixed point $\rho=\frac{1}{27}$.
The radial variable $\mathcal{R}$ in Eq. (12) of Ref. \cite{Reichenbach2006}, which gives the distance to the coexistence fixed point, is related to our $\rho$ by
\beq
\nonumber
\mathcal{R}^2 = 3\left(\frac{1}{27}-\rho\right).
\eeq
Our SDE \eqref{eq:RPSSDEEffRho}, independently of the precise form of $\overline{D}$, predicts an exponential decay for $\overline{\rho}$:
\beq
\nonumber
\overline{\rho(t)} = \rho(0) e^{-3t}.
\eeq
Starting from the coexistence fixed point $\rho(0)=\frac{1}{27}$, we obtain for the mean $\overline{\mathcal{R}^2(t)}$:
\beq
\label{eq:MeanROfT}
\overline{\mathcal{R}^2(t)} = \frac{1}{9}\left(1-e^{-3t}\right).
\eeq
For small times, this gives $\overline{\mathcal{R}^2(t)} = \frac{1}{3}t + \mathcal{O}(t^2)$ which is just the expression in Eq. (30) of Ref. \cite{Reichenbach2006}. The full form of \eqref{eq:MeanROfT} is its correct extension to long times.

We can now go beyond this and compute the fluctuations of $\mathcal{R}^2$ near the coexistence fixed point. Rewriting our SDE \eqref{eq:RPSSDEEffRho} in terms of the variable $\mathcal{R}^2$, we get
\beq
\nonumber
d(\mathcal{R}^2) = -3 d\rho = \left(\frac{1}{3}-3\mathcal{R}^2\right)dt + \sqrt{\overline{D}\left(\frac{1}{27}-\frac{1}{3}\mathcal{R}^2\right)}\, dV.
\eeq
Near the coexistence fixed point $\mathcal{R}=0$, taking the leading order of the drift term and the asymptotics \eqref{eq:DEffCoex} of $\overline{D}$, we obtain
\beq
\nonumber
d(\mathcal{R}^2) = \frac{1}{3}dt + \sqrt{\frac{2}{3}\mathcal{R}^2}\,dV.
\eeq
This shows that for small times, the distribution of $\mathcal{R}^2$ is exponential:
\beq
\nonumber
P(\mathcal{R}^2,t)=\frac{3}{t}e^{-\frac{3 \mathcal{R}^2}{t}}.
\eeq
In particular, we obtain the variance
\beq
\nonumber
\overline{\mathcal{R}^4-\overline{\mathcal{R}^2}^2}=\frac{t}{9}.
\eeq
}

In total, with Eq. \eqref{eq:RPSSDEEffRho} and Eq. \eqref{eq:RPSFPEEffRho} we have provided a description of the extinction process in the rock-paper-scissors game as an effective one-dimensional stochastic process on the space of the deterministically conserved quantity $\rho$. This process has a linear drift term $-3\rho$ and a complicated multiplicative noise which we computed exactly and which is given by Eq. \eqref{eq:RPSDiffCoeff}. { The asymptotics of our results near the coexistence fixed point $\rho=\frac{1}{27}$ reproduce the known results in \cite{Reichenbach2006}.}

\subsection{Constant Noise Approximation}

In order to avoid the stochastic averaging procedure and the long computation leading to the complicated expression for the noise coefficient in Eq. \eqref{eq:RPSDiffCoeff}, one might be tempted to take Eq. \eqref{eq:RPSSDEEffRho} and simplify it by replacing the multiplicative noise by a constant, additive noise as a rough approximation.

In this section, we will perform this constant noise approximation and compute some observables analytically. In the next section \ref{sec:SimResults}, we will see that close to the boundaries of phase space, numerical results deviate significantly from such a constant noise approximation. This shows that the computation of the nontrivial form of the diffusion coefficient in Eq. \eqref{eq:RPSDiffCoeff} is essential in order to obtain quantitatively correct results, especially close to the phase space boundaries.

Replacing the complicated function $\overline{D}(\rho)$ in Eq. \eqref{eq:RPSDiffCoeff} by a constant $D_0$, the extinction process in Eq.\eqref{eq:RPSSDEEffRho} is reduced to a standard \textit{Ornstein-Uhlenbeck process}:
\beq
d\rho = -\frac{3}{N}\rho\,dt + \frac{1}{\sqrt{N}}D_0\,dV .
\eeq
In this approximation, the dependence of the mean extinction time on the starting value of $\rho$ can be computed analytically in terms of the generalized hypergeometric function $\,_2F_2$:
\beqn
\nonumber
\frac{T_{\mathrm{ext}}(\rho)}{N} &=& \frac{C}{6}\sqrt{3 \pi D_0 } \mathrm{erfi}\left(\sqrt{\frac{3}{D_0}}\rho \right)  \\
& &-\frac{\rho^2}{D_0} \, _2F_2\left(1,1;\frac{3}{2},2;\frac{3 \rho^2}{D_0}\right) .
\label{eq:RPSConstNoiseExtTime}
\eeqn
$C$ is a constant fixed by the appropriate boundary conditions. From the singular nature of the boundary at $\rho_{\mathrm{max}}=\frac{1}{27}$ in Eq. \eqref{eq:RPSFPEEffRho} and Eq. \eqref{eq:RPSDiffCoeff}, one can derive that the mean extinction time must satisfy the boundary condition $T'_{\mathrm{ext}}(\rho_{\mathrm{max}})=-9$. In the constant noise approximation, this fixes the constant $C$ to be:
\begin{equation*}
C=\frac{1}{\rho_{\mathrm{max}}} e^{-\frac{3 \rho_{\mathrm{max}}^2}{D_0}}\left(\frac{1}{3}+\frac{3D_0}{2}\right) +\sqrt{\frac{\pi}{3 D_0}} \mathrm{erf}\left( \sqrt{\frac{3}{D_0}} \rho_{\mathrm{max}} \right).
\end{equation*}

The full extinction probability distribution depending on time, starting from a fixed initial condition $\rho_0$, can also be written down explicitely if the boundary at $\rho_{\mathrm{max}}$ is neglected. It then reads:
\beq
\label{eq:RPSConstNoiseSurvProb}
P_{\mathrm{ext}}(t,\rho_0)=\mathrm{erfc}\left( \rho_0 \sqrt{\frac{3}{D_0\left(e^{6t}-1\right)}}\right) .
\eeq
Asymptotically, this probability distribution possesses an exponential tail, $P_{\mathrm{surv}}(t) \propto e^{-3t}$, independent of the precise value of $D_0$. The exponent $-3t$ coincides very well with previous numerical results obtained in Ref. \cite{Ifti2003}.

\subsection{Comparison to simulations}
\label{sec:SimResults}

\begin{figure}[t]
\includegraphics[height=0.85\columnwidth,angle=-90]{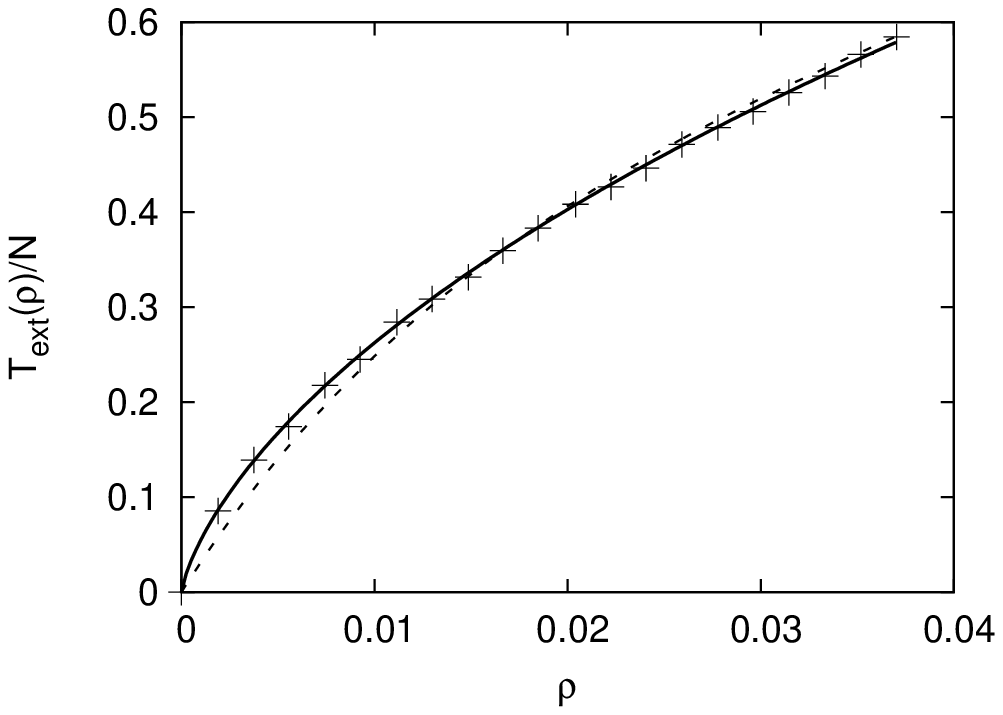}
\includegraphics[height=0.85\columnwidth,angle=-90]{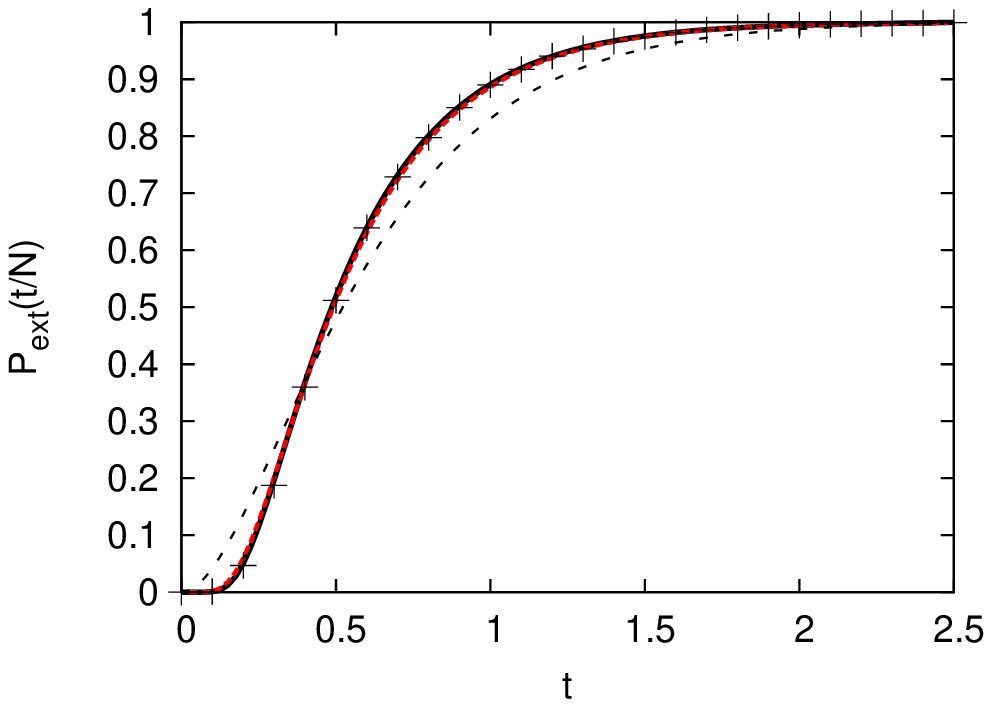}
\caption{{(Color online)} Comparison of theory and simulation results for the rock-paper-scissors game.
\textit{Top}: Mean extinction times depending on the initial condition. Solid curve: prediction obtained numerically from the stochastic averaging result in Eq. \eqref{eq:RPSSDEEffRho}. Dashed curve: Constant noise approximation, Eq. \eqref{eq:RPSConstNoiseExtTime}. Crosses: results of direct simulation of the reaction system in Eq. \eqref{eq:RPSReactionEqns} using the Gillespie algorithm for $N=1500$, averaged over $10^4$ realizations.
\textit{Bottom}: Extinction probability distribution, starting from the coexistence fixed point $\rho=\frac{1}{27}$ at $t=0$. Solid curve: prediction obtained numerically from the stochastic averaging result in Eq. \eqref{eq:RPSSDEEffRho}. Red {(thick)} dashed curve {(on top of solid black curve)}: Constant noise approximation, Eq. \eqref{eq:RPSConstNoiseSurvProb}. Black dashed curve: Phenomenological approximation previously proposed in Ref. \cite{Reichenbach2006}. Crosses: results of direct simulation of the reaction system in Eq. \eqref{eq:RPSReactionEqns} using the Gillespie algorithm for $N=3000$, averaged over $10^4$ realizations.}
\label{fig:PlotRPSExtProb}
\end{figure}

To verify the accuracy of the stochastic averaging procedure and the precise noise structure in Eq. \eqref{eq:RPSSDEEffRho}, we simulated the mean extinction times and the extinction probability distribution of the original reaction system in Eq. \eqref{eq:RPSReactionEqns} using an efficient algorithm due to Gillespie \cite{Gillespie1976,Gillespie1977}. The results are shown as crosses in figure \ref{fig:PlotRPSExtProb}.

This can then be compared to the predictions of the effective Fokker-Planck equation \eqref{eq:RPSFPEEffRho} with the full form of $\overline{D}(\rho)$ in Eq. \eqref{eq:RPSSDEEffRhoD} obtained by stochastic averaging. Although this effective Fokker-Planck equation cannot be solved analytically, the mean extinction times and survival probabilities can easily be determined numerically from the corresponding backward Fokker-Planck equation \cite{Gardiner}. The results are shown as blue lines in Fig.~\ref{fig:PlotRPSExtProb}. One can observe that the agreement to the simulation of the original reaction system in Eq.~\eqref{eq:RPSReactionEqns} is excellent.

Furthermore, Fig.~\ref{fig:PlotRPSExtProb} shows the purely analytical results of the constant noise approximation in the previous section (namely, Eqs.~\eqref{eq:RPSConstNoiseExtTime} and \eqref{eq:RPSConstNoiseSurvProb}) for $D_0 = 0.001$. Close to the {extinction }boundary {$\rho=0$, there is qualitative but no quantitative agreement between simulations and the constant noise approximation} (which is not surprising, considering the shape of $\overline{D}(\rho)$ close to $\rho=0$). This shows that close to the boundary, the precise form of the multiplicative noise plays a significant role, and provides further evidence for the correctness of Eq. \eqref{eq:RPSSDEEffRhoD}. {However, we find it interesting that when starting from the coexistence fixed point $\rho=\frac{1}{27}$, the constant noise approximation is in good quantitative agreement with simulations of both the mean extinction time and the extinction probability. This is surprising, since before going extinct the system will have to pass near the boundary $\rho=0$, where the approximation fails.}

Having given an extensive treatment of the cyclic three-species model, we will now increase the number of species by one and consider the four-species model with cyclic dominance.

\section{Cyclic Four-Species Model\label{sec:4Sp}}

In this section, we shall apply the formalism developed above to another example. We will consider the cyclic four-species Lotka-Volterra model, which is a natural object to study after the three-species rock-paper-scissors game discussed in the previous section. The reaction equations are given by:
\beqn
\nonumber A+B & \rightarrow A+A , \\
\nonumber B+C & \rightarrow B+B , \\
\nonumber C+D & \rightarrow C+C , \\
\label{eq:4SpReactionEqns} D+A & \rightarrow D+D .
\eeqn

For the case of equal reaction rates, the drift and noise terms for the Fokker-Planck equation \eqref{eq:GLVFPE} are obtained from Eq. \eqref{eq:GLVFPEAlpha} and Eq. \eqref{eq:GLVFPEB} as:
\beqn
\label{eq:4SpFPEAlpha}
& & \vec{\alpha} =\left(\begin{array}{c}
	a(b-d) \\
	b(c-a) \\
	c(d-b) \\
	d(a-c)
\end{array}\right) , \\
\label{eq:4SpFPEB}
& & B=\left(\begin{array}{cccc}
	a(b+d) & - a b & 0 & -a d \\
	- a b & b( c+ a) & - b c & 0 \\
	0 & -b c & c(d+b) & -cd \\
	-ad & 0 & -cd & d(a+c)
\end{array}\right) .
\eeqn

\subsection{Rate Equations\label{sec:4SpRE}}
The rate equations for this model can be written down directly from the drift term in Eq. \eqref{eq:4SpFPEAlpha}, and read:
\beq
\label{eq:4SpRateEqns}
\partial_t x_i = \alpha_i \quad \Leftrightarrow \quad
\left(\begin{array}{c}
	\partial_t a \\
	\partial_t b \\
	\partial_t c \\
	\partial_t d
\end{array}\right)
=\left(\begin{array}{c}
	a( b- d) \\
	b( c- a) \\
	c( d- b) \\
	d( a- c)
\end{array}\right) .
\eeq

\begin{figure}[t]
\centering
\includegraphics[width=0.75\columnwidth]{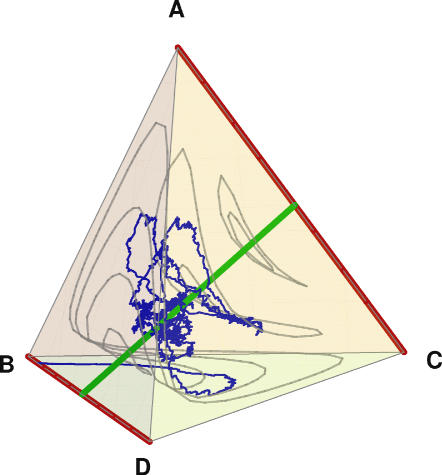}
\caption{{(Color online)} Phase space of the cyclic four-species model. Green {(light grey), diagonal}: Line of coexistence fixed points, Eq. \eqref{eq:4SpCoexLine}. Red {(dark grey) edges AC, BD}: Lines of extinction fixed points given in Eqns. \eqref{eq:4SpEdgeAC}, \eqref{eq:4SpEdgeBD}, where only two non-interacting species remain. Grey: Closed deterministic orbits for various values of $\tau_1$, $\tau_2$. Blue {(or dark grey)}: Sample stochastic trajectory for $N=300$.}
\label{fig:4SpPhaseSpace}
\end{figure}

As expected, the equations \eqref{eq:4SpRateEqns} keep the normalization condition $a+b+c+d=1$ invariant. The phase space $\{a,b,c,d|a,b,c,d \geq 0;a+b+c+d=1\}$ is now a three-dimensional simplex (i.e. a regular tetrahedron). Again, vertices correspond to extinction of all but one species, edges correspond to extinction of two out of the four species, and faces to extinction of one species.

The fixed points of the rate equations \eqref{eq:4SpRateEqns} form three lines in the phase space simplex:
\begin{itemize}
	\item One line of coexistence fixed points given by $b=d$ and $a=c$, i.e. parametrized by 
	\beq
	\label{eq:4SpCoexLine}
	(a,b,c,d)=\left(t,\frac{1}{2}-t,t,\frac{1}{2}-t\right) \quad\mbox{with}\quad t\in \left[0;\frac{1}{2}\right] .
	\eeq
	\item The edge AC, i.e. all states with coexistence of A and C only, parametrized by 
	\beq
	\label{eq:4SpEdgeAC}
	(a,b,c,d)=(t,0,1-t,0) \quad\mbox{with}\quad t\in [0;1] .
	\eeq
	\item The edge BD, i.e. all states with coexistence of B and D only, parametrized by 
	\beq
	\label{eq:4SpEdgeBD}
	(a,b,c,d)=(0,t,0,1-t) \quad\mbox{with}\quad t\in [0;1] .
	\eeq
\end{itemize}
A graphical representation of the structure of the fixed points is shown in figure \ref{fig:4SpPhaseSpace}.

It is straightforward to check that the trajectories solving the equations \eqref{eq:4SpRateEqns} now exhibit two conserved quantities:
\beqn
\nonumber \tau_1 & = &ac , \\
\tau_2 & = & bd .
\eeqn

The curves given by $\tau_1=const.$, $\tau_2=const.$ are closed, neutrally stable orbits around the line of coexistence fixed points. Close to this line, they are almost circular, while further away they approach the shape of the simplex boundaries. A few exemplary orbits are shown in figure \ref{fig:4SpPhaseSpace}.

In total, the deterministic dynamics for the four-species cyclic Lotka-Volterra model is quite similar to the behaviour in the three-species case. The relative abundances of the species oscillate indefinitely along a fixed, closed orbit in phase space, and no extinction takes place.

\subsection{Stochastic Extinction Process}
We would now like to investigate the behaviour of the four-species model when stochastic fluctuations, modelled by the noise terms in Eq. \eqref{eq:GLVSDE}, are included. According to the experience from the three-species rock-paper-scissors model, we again expect to see extinction on a time scale $\propto N$ since the deterministic orbits are neutrally stable.

However, as already apparent from the description of the rate equation dynamics in the previous section, the set of fixed points is now much larger than in the rock-paper-scissors game.  Inserting the parametrizations of the fixed lines in equations \eqref{eq:4SpCoexLine}, \eqref{eq:4SpEdgeAC}, \eqref{eq:4SpEdgeBD} into \eqref{eq:4SpFPEB}, we see that the noise matrix for the Fokker-Planck equation vanishes on the edges AC and BD, but not on the line of coexistence fixed points.

Hence, the absorbing states for the stochastic process are precisely the edges AC and BD of the phase space simplex, parametrized by the equations \eqref{eq:4SpEdgeAC} and \eqref{eq:4SpEdgeBD}, corresponding to states with a mixture of non-interacting species (A and C or B and D). This is consistent with the picture obtained directly from the reaction equations \eqref{eq:4SpReactionEqns}, where it is clear that these are exactly the states where no more reactions can occur.

In order to analyze the stochastic model quantitatively, we now pursue the same approach used for the rock-paper-scissors model and investigate the behaviour of the deterministically conserved quantities when stochasticity is included.
Applying the It\^{o} formula and using the Langevin equations \eqref{eq:GLVSDE}, we obtain the following stochastic differential equation for the conserved quantities {$\tau_\mu$, $\mu=1,2$}:
\beq
\label{eq:4SpSDETau}
d\tau_{\mu} = \frac{1}{\sqrt{N}}\sum_{i,j}\left(\mathcal{C}_{ij}\partial_i\tau_{\mu}\right)\, dW_j .
\eeq
Note that in contrast to the corresponding calculation in \ref{sec:RPSStochAvg}, there are no stochastic drift terms due to the specific form of $B$ in Eq. \eqref{eq:4SpFPEB}. Just as for the noise in the 3-species case, the prefactor of the noises in Eq. \eqref{eq:4SpSDETau} still depends on the individual species concentrations $a$, $b$, $c$, $d$ which vary along a trajectory with fixed $\tau_1$, $\tau_2$.

Stochastic averaging now again permits us to obtain a closed system of equations describing the two-dimensional slow process $(d\tau_1 (t), d\tau_2 (t))$:
\beq
\label{eq:4SpSDETaus2}
d\tau_{\mu} = \frac{1}{\sqrt{N}}\sum_{{\nu}=1}^2\mathcal{D}_{\mu\nu}\;dV_{\nu} ,
\eeq
As previously, $\tau_1$ and $\tau_2$ are now treated as free variables. $dV_{\nu}$, ${\nu}=1,2$ are two independent Gaussian noises. The region of $\tau_1$-$\tau_2$ space on which the process given by Eq. \eqref{eq:4SpSDETaus2} occurs is bounded by the absorbing boundaries $\tau_1=0$ and $\tau_2=0$, as well as a reflecting boundary at 
\beq
\label{eq:4SpCurvedBdry}
\sqrt{\tau_1}+\sqrt{\tau_2}=\frac{1}{2}.
\eeq
The matrix of the effective diffusion coefficients $\mathcal{D}$ is defined by $\mathcal{D} \mathcal{D}^T=\overline{B_\tau}$, where $\overline{B_\tau}$ is the orbit average of the correlation matrix in Eq. \eqref{eq:4SpSDETau}:
\beqn
\nonumber \overline{B_\tau} & = & N
\left(\begin{array}{cc}
	\overline{\left\langle d\tau_1 d\tau_1 \right\rangle} & \overline{\left\langle d\tau_1 d\tau_2 \right\rangle} \\
	\overline{\left\langle d\tau_2 d\tau_1 \right\rangle} & \overline{\left\langle d\tau_2 d\tau_2 \right\rangle}
\end{array}\right) \\
\nonumber
 & =&
\left(\begin{array}{cc}
	\overline{(\nabla\tau_1)B(\nabla\tau_1)} & \overline{(\nabla\tau_1)B(\nabla\tau_2)} \\
	\overline{(\nabla\tau_2)B(\nabla\tau_1)} & \overline{(\nabla\tau_2)B(\nabla\tau_2)}
\end{array}\right) \\
\nonumber
 & = &
\left(\begin{array}{cc}
	\tau_1 \overline{(a+c)(b+d)} & -4\tau_1 \tau_2 \\
	-4\tau_1 \tau_2 & \tau_2 \overline{(a+c)(b+d)}
\end{array}\right) \\
\label{eq:4SpNoiseTaus}
 &=:&
\left(\begin{array}{cc}
	\tau_1 f(\tau_1,\tau_2) & -4\tau_1 \tau_2 \\
	-4\tau_1 \tau_2 & \tau_2 f(\tau_1,\tau_2)
\end{array}\right) .
\eeqn

The high degree of symmetry and the simplicity of the model allow one to give an analytic expression for the function $f$, providing an exact expression for Eq. \eqref{eq:4SpNoiseTaus}:
\beq
\label{eq:4SpNoiseF}
f(\tau_1,\tau_2):=\overline{ (a+c)(b+d)}=h(\tau_1,\tau_2)\frac{E\left(k(\tau_1,\tau_2)\right)}{K\left(k(\tau_1,\tau_2)\right)}
\eeq
Again, $K(k)$ and $E(k)$ are complete elliptic integrals of the first and second kind, respectively. The elliptic modulus and the helper function $h$ are given by:
\beqn
\label{eq:4SpKinTau}
&&k^2(\tau_1,\tau_2)= 4\frac{\sqrt{\sigma_1^2-\sigma_2}}{h(\tau_1,\tau_2)} , \\
\label{eq:4SpHinTau} 
&&h(\tau_1,\tau_2) = \frac{1}{2} \left( \sigma_1+\sqrt{\sigma_1^2-\sigma_2} \right) ,
\eeqn
with
\beq
\nonumber
\sigma_1 = 1-4\tau_1-4\tau_2\quad\quad\sigma_2=64\tau_1\tau_2 .
\eeq
For details of the calculation leading to these expressions, see appendix \ref{sec:AppCalc4Sp}.

Equivalently to Eq. \eqref{eq:4SpSDETaus2}, we can write the effective stochastic process as a Fokker-Planck equation:
\beq
\label{eq:4SpFPETaus}
\partial_t P(\tau_1,\tau_2,t) = \frac{1}{N}\sum_{{\mu,\nu}=1}^2 \partial_{\mu} \partial_{\nu} \left[ (\overline{B_\tau})_{\mu\nu} P(\tau_1,\tau_2,t) \right] .
\eeq

Note how this calculation provides a natural generalization of the analysis performed for the rock-paper-scissors game (section \ref{sec:RPSStochAvg}). As is apparent from Eq. \eqref{eq:4SpSDETaus2} (or its Fokker-Planck equivalent Eq. \eqref{eq:4SpFPETaus}), we again obtain a dynamics for the variables $\tau$ which occurs on a time scale $\propto N$.

In total, we have obtained a complete description of the extinction process in the four-species cyclic model as a two-dimensional diffusion process with varying diffusion coefficients. 

\subsection{Comparison to simulations}

\begin{figure}[t]
\includegraphics[height=0.85\columnwidth,angle=-90]{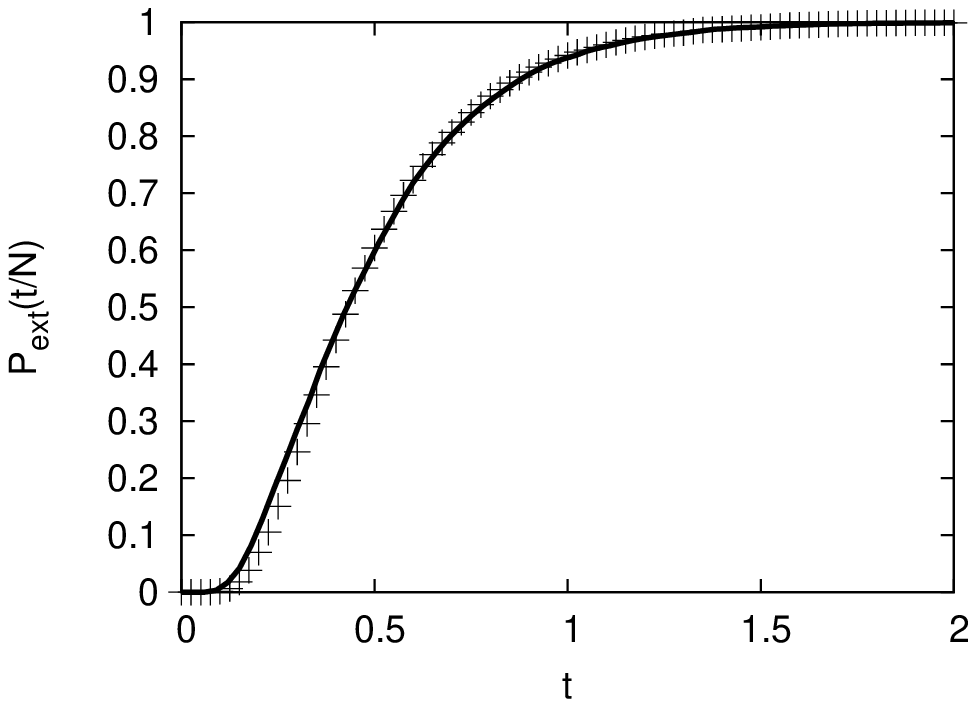}
\includegraphics[height=0.85\columnwidth,angle=-90]{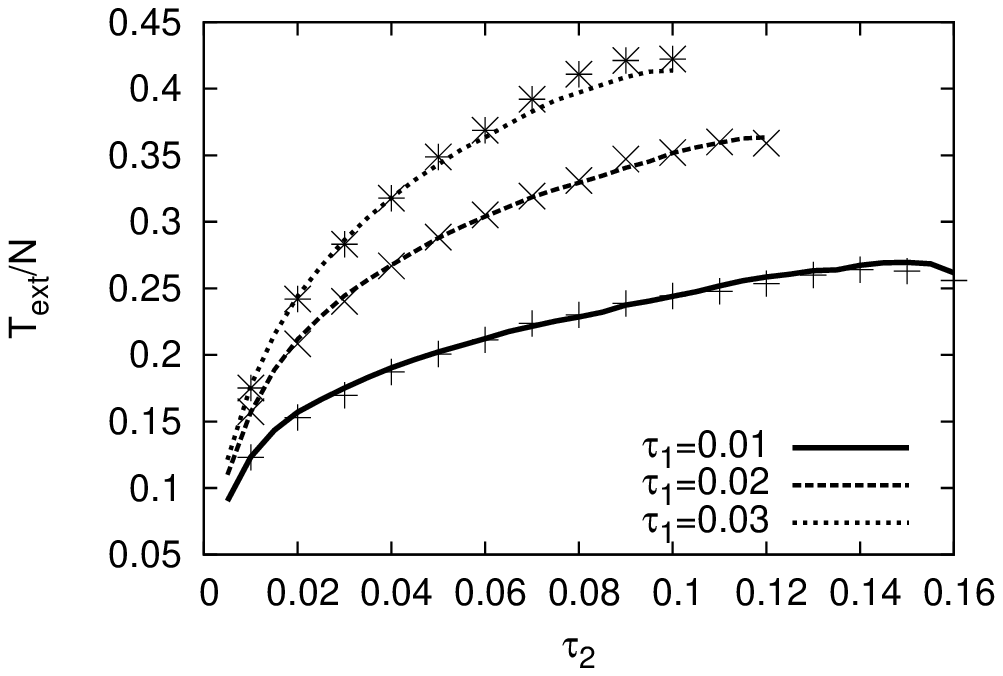}
\caption{Comparison of theory and simulation results for the four-species model of cyclic dominance.
\textit{Top}: Extinction probability distribution, starting from $a=b=c=d=\frac{1}{4}$ at $t=0$. Solid curve: Simulation of effective stochastic process in Eq. \eqref{eq:4SpSDETaus2}, averaged over $10^4$ realizations. Crosses: Gillespie simulation of the reaction system in Eq. \eqref{eq:4SpReactionEqns} with a system size $N=8000$, averaged over $10^4$ realizations.
\textit{Bottom}: Mean extinction times depending on initial conditions. Lines: Simulations of effective stochastic process in Eq. \eqref{eq:4SpSDETaus2}, averaged over $10^4$ realizations. Crosses: Gillespie simulations of the reaction system in Eq. \eqref{eq:4SpReactionEqns} with a system size $N=2000$, averaged over $10^3$ realizations.}
\label{fig:Plot4SpCyclicExtTime}
\end{figure}

As for the rock-paper-scissors game, we can now verify the accuracy with which various quantities of interest for the extinction process can be predicted by the effective stochastic process in Eq. \eqref{eq:4SpSDETaus2}. Since this is now a two-dimensional stochastic process in a region with a complicated shape and mixed boundary conditions, it is much harder to treat than the one-dimensional effective process in the three-species case. 

Determining the mean extinction times and extinction probabilities from the effective Fokker-Planck equation \eqref{eq:4SpSDETaus2} as was done in the three-species case is not feasible here, since it would require solving elliptic second-order PDE's over a domain bounded by Eq. \eqref{eq:4SpCurvedBdry}. Instead, we obtained mean extinction times and extinction probabilities from stochastic simulations of the effective Langevin equations \eqref{eq:4SpSDETaus2} using the XmdS package \cite{Hope}.

The results are shown in figure \ref{fig:Plot4SpCyclicExtTime}. It can be observed that the predictions of the effective Langevin equations \eqref{eq:4SpSDETaus2} compare very well to the results of direct simulation of the original reaction system in Eq. \eqref{eq:4SpReactionEqns}. We attribute the slight discrepancies in mean extinction times close to the boundary given by Eq. \eqref{eq:4SpCurvedBdry} to the general difficulty of simulating a stochastic process near a curved reflecting boundary. The stochastic averaging method becomes exact at this boundary, since the deterministic orbits are the individual coexistence fixed points.

\section{Conclusion\label{sec:Conclusion}}

We have analyzed the extinction process in two stochastic Lotka-Volterra models which are neutrally stable in the deterministic formulation. We have seen that when fluctuations are included, the deterministically conserved quantities change slowly (on a time scale proportional to the population size) and drive extinction. After some finite time, only non-interacting species remain.

The separation of time scales between the rapid oscillations described by the deterministic rate equations and the slow movement between different orbits due to noise allowed us to apply the method of stochastic averaging. By doing this, we removed the fast, oscillatory degrees of freedom and gave a quantitative description of the extinction process using effective stochastic differential equations on the space of deterministically conserved quantities.

We have obtained various quantities of interest for the extinction process from these effective equations, and observed that they agree very well with direct simulations.

The stochastic averaging procedure required computation of certain integrals over the closed deterministic orbits, which we were able to perform analytically in the toy models we considered. In more complicated models with less symmetry (e.g. different reaction rates), this may not be possible anymore. However, even for complicated and asymmetric closed deterministic orbits, the averaging required to determine the effective drift and diffusion coefficients can easily be performed numerically. Thus, we think that our approach should be just as helpful for elucidating the impact of noise in more general models.

As we saw, a considerable advantage of the stochastic averaging method (especially in comparison to the treatment in Ref. \cite{Reichenbach2006}) is that neither the drift nor the noise terms need to be linearized. The full, nonlinear dynamics of the model and the multiplicative noise structure, as well as the complex geometry of the phase space can be taken into account. Furthermore, it is not necessary to write the dynamics explicitely in terms of a radial and a phase variable. This is useful since e.g. in the rock-paper-scissors model in section \ref{sec:RPS}, there is no obvious choice for a canonical phase variable. 

The idea of describing the extinction process by the evolution of a deterministically conserved variable was also utilized by Parker and Kamenev in Ref. \cite{Parker2009}. They applied it in a semiclassical approximation and obtained the asymptotics of the extinction probability distribution in the standard two-species Lotka-Volterra model. However, our approach is quite different technically and allows a straightforward generalization to more complex models containing more than two species (as in the example in section \ref{sec:4Sp}).

We also expect that it should be possible to extend this treatment away from the borderline case of neutral stability to weakly stable or unstable models. Heuristically, this would give rise to a deterministic drift term in equation \eqref{eq:RPSSDEEffRho} (or its analogues) which is independent of $N$ but controlled by some other small expansion parameter. Investigating how such models can be constructed in a natural way and the details of this generalization requires further research. Indeed, for a recent study along these lines see Refs. \cite{Case2010,Durney2011}.

It is also important to understand if and how the present method can be extended to models with spatial degrees of freedom, as discussed in Refs. \cite{He2010,Tauber2011}. They exhibit much more complex phenomena (see e.g. \cite{Reichenbach2007, Szabo2007, Perc2007, Frey2010,Tauber2011}) and are more realistic than the well-mixed models we discuss here.

Another approach to investigating the effects of stochasticity on similar models was pursued in Refs. \cite{Alonso2007,Dauxois2009,McKane2005,Boland2009}, where the effects of the stochasticity on the phase variable and the spectral distribution of its oscillation were investigated. This is complementary to our treatment, where we focus on the radial variables instead. This allows us to capture the dynamics of the extinction process.

In a more general context, the present discussion gives an illustration of how stochasticity may change the behaviour in a nonlinear dynamical system qualitatively by adding a stochastic drift to a deterministically conserved quantity. It also provides some ideas for treating the effects of complicated, multiplicative noise on such systems analytically. This may be of considerable interest for non-equilibrium statistical physics in general.

\acknowledgements
We would like to thank T. Reichenbach and J. Cremer for helpful discussions. Financial support of the German Research Foundation via the SFB TR12 ``Symmetries and Universalities in Mesoscopic Systems'' is gratefully acknowledged. 

\appendix
\section{Computation of noise term for three-species cyclic model\label{sec:AppCalc3Sp}}
In this appendix, we shall sketch the computation of the stochastic averaging integral in Eq. \eqref{eq:RPSSDEEffRhoD}.

We will parametrize an orbit with fixed $\rho$ by one of the species' concentrations, e.g. $a$.
As is easily verified using the method of Lagrange multipliers, the extremal values $a_{\mathrm{min}}$ and $a_{\mathrm{max}}$ which $a$ assumes on such an orbit are real roots of the polynomial
\beq
a(1-a)^2=4\rho.
\eeq
The third root of this polynomial is then also real, and will be denoted by $a_1$. As is well known, explicit expressions for all three roots exist.

We hence write down the factorization
\beq
\label{eq:RPSEqRatesExtrA2}
a(1-a)^2-4\rho = (a-a_{\mathrm{min}})(a-a_{\mathrm{max}})(a-a_1) .
\eeq

Now, let us give an explicit parametrization of each orbit. We will choose $\rho$ and $a$ as the independent variables, with $0<\rho<\frac{1}{27}$ and $a_{\mathrm{min}}<a<a_{\mathrm{max}}$. Then $b$ and $c$ are given by
\beqn
\label{eq:RPSEqRatesMFb} b_{1,2} & = & \frac{a(1-a) \pm \sqrt{a^2 (1-a)^2-4a\rho}}{2a} , \\
\label{eq:RPSEqRatesMFc} c_{1,2} & = & \frac{a(1-a) \mp \sqrt{a^2 (1-a)^2-4a\rho}}{2a} .
\eeqn
In each case, the + respectively - signs correspond to the two branches of the orbit for a fixed value of $a$.

Inserting Eq. \eqref{eq:RPSEqRatesMFb} and Eq. \eqref{eq:RPSEqRatesMFc} into the rate equation $\partial_t a = a (b-c)$ we get:
\beq
\label{eq:RPSEqRatesAdot}
\partial_t a = \pm \sqrt{a^2 (1-a)^2-4a\rho} .
\eeq

Now, let us calculate the period of an orbit, $T(\rho)$. By a simple substitution we have
\beq
T(\rho)=\int_0^{T(\rho)}{1\,dt}=2\int_{a_{\mathrm{min}}}^{a_{\mathrm{max}}}{\frac{da}{\partial_t a}} .
\eeq
The factor $2$ arises since each orbit has two symmetric branches, when parametrized by e.g. $a$. 
Inserting Eq. \eqref{eq:RPSEqRatesAdot} and Eq. \eqref{eq:RPSEqRatesExtrA2}, we get 
\beq
T(\rho)=2\int_{a_{\mathrm{min}}}^{a_{\mathrm{max}}}{\frac{da}{\sqrt{a(a-a_{\mathrm{min}})(a-a_{\mathrm{max}})(a-a_1)}}} .
\eeq

This is a standard integral that can be expressed in terms of $K(k)$, the complete elliptic integral of the first kind (see e.g. \cite{Byrd1971}):
\beq
\label{eq:RPSEqRatesPeriod}
T(\rho)=\frac{4 K(k)}{\sqrt{(a_1-a_{\mathrm{min}})a_{\mathrm{max}}}} .\\
\eeq
The elliptic modulus $k$ is given by Eq. \eqref{eq:RPSEqRatesModulus}.

The last remaining piece we need is the average of $\frac{1}{a}$ over a deterministic orbit. Again applying a substitution and using Eq. \eqref{eq:RPSEqRatesAdot} and Eq. \eqref{eq:RPSEqRatesExtrA2}, we have:
\beqn
\nonumber
& & \int_0^{T(\rho)}{\frac{dt}{a}}  =  2\int_{a_{\mathrm{min}}}^{a_{\mathrm{max}}}{\frac{da}{a\, \partial_t a}}\\
\nonumber
& & = 2\int_{a_{\mathrm{min}}}^{a_{\mathrm{max}}}{\frac{da}{a \sqrt{a(a-a_{\mathrm{min}})(a-a_{\mathrm{max}})(a-a_1)}}} .
\eeqn

This, too, is a standard integral that can be expressed in terms of complete elliptic integrals (see e.g. \cite{Byrd1971}), giving:
\beqn
\nonumber
\int_0^{T(\rho)}{\frac{dt}{a}} & = & \frac{4\left[ (a_1-a_{\mathrm{max}}) \Pi (k^2, k) +a_{\mathrm{max}}K(k)  \right]}{a_1 a_{\mathrm{max}}\sqrt{(a_1-a_{\mathrm{min}})a_{\mathrm{max}}} } \\
\nonumber
 & = & \frac{4 \sqrt{(a_1-a_{\mathrm{min}})a_{\mathrm{max}}}}{a_1 a_{\mathrm{max}} a_{\mathrm{min}}} E(k)  \\
& & 
 + \frac{4\,K(k)}{a_1 \sqrt{(a_1-a_{\mathrm{min}})a_{\mathrm{max}}}} .
 \label{eq:RPSEqRatesOneOverAAvg}
\eeqn
Here, $E(k)$ is the complete elliptic integral of the second kind, $\Pi(n,k)$ is the complete elliptic integral of the third kind, and the elliptic modulus $k$ is again given by Eq. \eqref{eq:RPSEqRatesModulus}. The second line in Eq. \eqref{eq:RPSEqRatesOneOverAAvg} was obtained by applying the relation $\Pi(k^2,k)=\frac{E(k)}{1-k^2}$.

Combining Eq. \eqref{eq:RPSEqRatesOneOverAAvg} and Eq. \eqref{eq:RPSEqRatesPeriod} we get the result used in Eq. \eqref{eq:RPSDiffCoeff}.

\section{Computation of noise term for four-species cyclic model\label{sec:AppCalc4Sp}}
The computation of the average of the correlation matrix for the four-species model, Eq. \eqref{eq:4SpNoiseTaus}, works along the same lines as the three-species case in appendix \ref{sec:AppCalc3Sp}.

We parametrize the deterministic orbit with fixed $\tau_1$, $\tau_2$ in terms of $a$. The extremal values of $a$ are given as:
\beq
\label{eq:4SpAminmax}
a_{\mathrm{max,min}}=\frac{1}{2}\left( 1-2\sqrt{\tau_2} \pm \sqrt{(1-2\sqrt{\tau_2})^2 -4\tau_1} \right).
\eeq
The other variables are expressed in terms of $a$, $\tau_1$ and $\tau_2$ as:
\beqn
\nonumber b_{1,2} & = & \frac{1}{2}\left(1-a-\frac{\tau_1}{a} \pm \sqrt{\left(1-a-\frac{\tau_1}{a}\right)^2-4\tau_2}\right) , \\ 
\nonumber d_{1,2} &= &\frac{1}{2}\left(1-a-\frac{\tau_1}{a} \mp \sqrt{\left(1-a-\frac{\tau_1}{a}\right)^2-4\tau_2}\right) , \\
\label{eq:4SpOrbitParam} c & = & \frac{\tau_1}{a} .
\eeqn

As for the orbits seen in the three-species case, the plus and minus signs correspond to the two (symmetrical) branches of an orbit for each $a$.
From Eq. \eqref{eq:4SpOrbitParam} we get:
\beq
\label{eq:4SpOrbitAdot}
\partial_t a=a(b-d)=\pm\sqrt{(a(1-a)-\tau_1)^2-4a^2\tau_2} .
\eeq

With all of these results, we can now calculate the period of an orbit with fixed $\tau_1$, $\tau_2$:
\beqn
T(\tau_1,\tau_2)&=& 2\int_{a_{\mathrm{min}}}^{a_{\mathrm{max}}}{\frac{da}{\sqrt{(a(1-a)-\tau_1)^2-4a^2\tau_2}}} .
\label{eq:4SpOrbitPeriod1}
\eeqn
Again, the factor $2$ arises from the two symmetric branches of each orbit. 
The denominator of the integrand can now be factored as:
\beqn
\nonumber
& & \sqrt{(a(1-a)-\tau_1)^2-4a^2\tau_2} \\
& & =\sqrt{(a_{\mathrm{max}}-a)(a-a_{\mathrm{min}})(a_1-a)(a-a_2)},
\label{eq:4SpAfactor2}
\eeqn
where 
\beq
\label{eq:4SpA12}
a_{1,2}=\frac{1}{2}\left( 1+2\sqrt{\tau_2} \pm \sqrt{(1+2\sqrt{\tau_2})^2 -4\tau_1} \right) .
\eeq
Note that $a_1 > a_{\mathrm{max}}$ and $a_2 < a_{\mathrm{min}}$. Furthermore, Vieta's theorem gives following relations between $a_{\mathrm{min}}$, $a_{\mathrm{max}}$, $a_1$, $a_2$:
\beqn
\nonumber a_1 a_2 = a_{\mathrm{min}} a_{\mathrm{max}} = \tau_1 , \\
\nonumber a_1 + a_2 = 1+2\sqrt{\tau_2} ,\\
\label{eq:4SpVieta} a_{\mathrm{min}} + a_{\mathrm{max}} = 1-2\sqrt{\tau_2} .
\eeqn
These will be very useful for simplifying some expressions later on.

Upon inserting Eq. \eqref{eq:4SpAfactor2} into the integral in Eq. \eqref{eq:4SpOrbitPeriod1}, we get a standard elliptic integral 
\beq
T(\tau_1,\tau_2)= \frac{4}{\sqrt{h(\tau_1,\tau_2)}} K\left(k(\tau_1,\tau_2)\right) .
\label{eq:4SpOrbitPeriod2}
\eeq
Here, $h$ is a helper function defined by
\beq
\label{eq:4SpHDef}
h(\tau_1,\tau_2):=(a_1-a_{\mathrm{min}})(a_{\mathrm{max}}-a_2) .
\eeq
$K(k)$ is the complete elliptic integral of the first kind, with the elliptic modulus given by
\beq
\label{eq:4SpEllipticModulus}
k^2=\frac{(a_{\mathrm{max}}-a_{\mathrm{min}})(a_1-a_2)}{(a_1-a_{\mathrm{min}})(a_{\mathrm{max}}-a_2)} .
\eeq
By using Eq. \eqref{eq:4SpVieta}, Eq. \eqref{eq:4SpAminmax} and Eq. \eqref{eq:4SpA12} we can express the helper function and the elliptic modulus explicitly in terms of $\tau_1$ and $\tau_2$, giving the expressions in Eq. \eqref{eq:4SpKinTau} and Eq. \eqref{eq:4SpHinTau}.
Observe that in Eq. \eqref{eq:4SpKinTau} and Eq. \eqref{eq:4SpHinTau}, the symmetry in $\tau_1$ and $\tau_2$ (which is required by the cyclic symmetry of the model, but was lost when we chose to parametrize the orbit explicitely using the variable $a$) is again manifest.

To calculate the average of the noise matrix $B_\tau$ we also need the following integral:
\beq
\overline{(a+c)(b+d)}= \frac{1}{T(\tau_1,\tau_2)}\int_0^{T} \left(a+\frac{\tau_1}{a}\right)\left(1-a-\frac{\tau_1}{a}\right)\,dt .
\label{eq:4SpStochAvgReq}
\eeq

This is reduced to the following four basic elliptic integrals:
\beqn
\label{eq:4SpStochAvgInt3}
\nonumber I_1 &=& \int_{a_{\mathrm{min}}}^{a_{\mathrm{max}}}{\frac{a\,da}{\sqrt{(a_1-a)(a_{\mathrm{max}}-a)(a-a_{\mathrm{min}})(a-a_2)}}} ,\\
\nonumber I_2 &=& \int_{a_{\mathrm{min}}}^{a_{\mathrm{max}}}{\frac{da}{a\sqrt{(a_1-a)(a_{\mathrm{max}}-a)(a-a_{\mathrm{min}})(a-a_2)}}} ,\\
\nonumber I_3 &=& \int_{a_{\mathrm{min}}}^{a_{\mathrm{max}}}{\frac{a^2\,da}{\sqrt{(a_1-a)(a_{\mathrm{max}}-a)(a-a_{\mathrm{min}})(a-a_2)}}} ,\\
\nonumber I_4 &=& \int_{a_{\mathrm{min}}}^{a_{\mathrm{max}}}{\frac{da}{a^2\sqrt{(a_1-a)(a_{\mathrm{max}}-a)(a-a_{\mathrm{min}})(a-a_2)}}}.
\eeqn
These can be computed using the formulae in Ref. \cite{Byrd1971} giving:
\beqn
\nonumber I_1 &=\frac{2}{\sqrt{h}} &\left[a_1 K(k) -(a_1-a_{\mathrm{max}}) \Pi({p},k)\right] ,\\
\nonumber I_2 &=\frac{2}{\sqrt{h}} &\left[\frac{1}{a_1} K(k) -\left(\frac{1}{a_1}-\frac{1}{a_{\mathrm{max}}}\right) \Pi({q},k)\right] ,\\
\nonumber I_3 &= \frac{2}{\sqrt{h}} &\left[a_1^2 K(k) -2 a_1 (a_1-a_{\mathrm{max}}) \Pi({p},k) \right.\\
\nonumber & & \left. +(a_1 - a_{\mathrm{max}})^2 V_2 ({p})\right] ,\\
\nonumber I_4 &=\frac{2}{\sqrt{h}} &\left[\frac{1}{a_1^2} K(k) - \frac{2}{a_1} \left(\frac{1}{a_1}-\frac{1}{a_{\mathrm{max}}}\right) \Pi({q},k) \right. \\
\nonumber & & \left.+\left(\frac{1}{a_1}-\frac{1}{a_{\mathrm{max}}}\right)^2 V_2({q})\right] .
\eeqn
As usual, $K(k)$, $E(k)$ and $\Pi(n,k)$ are the complete elliptic integrals of the first, second and third kinds, respectively. ${p}$, ${q}$ and $V_2$ are given in our notation by:
\beqn
\nonumber {p} & = & \frac{a_{\mathrm{max}}-a_{\mathrm{min}}}{a_1-a_{\mathrm{min}}},\\
\nonumber {q} & = & \frac{a_1(a_{\mathrm{max}}-a_{\mathrm{min}})}{a_{\mathrm{max}}(a_1-a_{\mathrm{min}})}.
\eeqn
$V_2$ is defined as
\beqn
\nonumber
V_2({x}) &=& \frac{1}{2({x}-1)(k^2-{x})} \left[ {x} E + (k^2-{x}) K \right. \\
\nonumber 
& & \left. +(2{x} k^2 +2{x}-{x}^2-3k^2)\Pi({x})\right]
\eeqn
Here, we dropped the elliptic modulus $k$ (which is always the same) from the arguments of $K$, $E$ and $\Pi$.
Combining the expressions for $I_1, ..., I_4$, we obtain after some long and tedious algebra the surprisingly simple result in Eq. \eqref{eq:4SpNoiseF} for Eq. \eqref{eq:4SpStochAvgReq}. This result, too, is symmetric in $\tau_1$ and $\tau_2$ (as expected, since the quantity $(a+c)(b+d)$ is invariant under cyclic permutations).


\end{document}